\documentclass[aps,floatfix,showpacs,twocolumn,superscriptaddress]{revtex4-1}
\usepackage{amssymb}
\usepackage{hyperref}
\usepackage{amsmath}
\usepackage{graphicx}
\usepackage{dsfont}
\usepackage{epstopdf}
\usepackage{amsfonts}
\usepackage{color}
\usepackage{hyperref}
\usepackage{bm}
\usepackage{array}
\usepackage{tabularx}
\usepackage{mathtools}
\usepackage{braket}
\usepackage{amsmath}
\usepackage{relsize}
\usepackage{filecontents}
\usepackage{soul}
\usepackage{natbib}
\usepackage{dsfont}

\begin{document}

\title{A $\pi$-shaped Quantum Device for Implementation of Bell States in Solid State Environment}
\author{Aman Ullah}
 \affiliation{Department of Physics, School of Natural Sciences (SNS), National University of Sciences and Technology (NUST), Islamabad 44000, Pakistan}
 \author{Mohammad Ali Mohammad}
\affiliation{School of Chemical and Materials Engineering (SCME), National University of Sciences and Technology, Islamabad 44000, Pakistan}
\author{Mahmood Irtiza Hussain }
\affiliation{Institute for Quantum optics and Quantum Information
Technikerstr. 21a
6020 Innsbruck, Austria}
\author{Syed Rizwan}
\email{Corresponding author: Syed Rizwan\\ Email: syedrizwanh83@gmail.com}
\affiliation{Department of Physics, School of Natural Sciences, National University of Sciences and Technology (NUST), Islamabad 44000, Pakistan}

\begin{abstract}
\section{abstract}
Electronic spin-qubit is key ingredient for quantum information processing in a solid state environment. We present a $\pi$-shaped two-qubit entanglement device capable of measuring the resultant states in Bell basis. In our device, source spins ($\uparrow_s$ or $\downarrow_s$) are electrically generated and tunnelled to channel where they interact with channel spins ($\uparrow_c$ or $\downarrow_c$) via exchange interaction which is responsible for 2-qubit entanglement. Electrical control over spins gives rise to the Bell states. The $U_{\sqrt{SWAP}}$ and CNOT gate operations are implemented through these Bell states for universal quantum computation. $\pi$-shaped quantum device can be used as a solid state interconnect between different parts of the circuit in integrated chips.

\keywords:{Keywords: Quantum entanglement; Rashba Spin-Orbit Interaction; Quasi Linearization Method; $U_{SWAP}$ operation}
\end{abstract}
\date{\today}\maketitle
\section{Introduction}
The electron spin degree of freedom is a known resource for quantum information processing (QIP) in a solid state environment \cite{tittel2000quantum,popescu2004all,kloeffel2013prospects,mohiyaddin2016transport,loss1998quantum}. Entanglement generation in multiple spin-qubits (SQu) and their manipulation in quantum device is quite challenging \cite{salter2010entangled,horodecki2009quantum,dickel2018chip}.
Several approaches have been adopted for entanglement generation between qubits of various forms including photons, trapped ions, cold atoms, spin and charge states in quantum dots (QDs), dopants in
solids to name several examples\cite{watson2018programmable,bohnet2016quantum,coe2010hubbard,knill2001scheme,ladd2010quantum,van2002electron,hayat2014cooper,homid2015efficient,hussain2014geometric}.

For large scale quantum computing, the coherence in the quantum states leaks out quickly \cite{klobus2014entanglement}. To preserve entanglement during QIP, photonic interconnects are being suggested for multiprocessing of qubits \cite{foxen2017qubit,wenner2017deterministic,wang2016chip,miller2009device,soref2006past}. The disadvantage of using photonic interconnects is the loss of fidelity in quantum states during the information processing between SQu and photonic qubit. To avoid loss of fidelity, solid state interconnects using QDs have been suggested \cite{saraga2003spin,legel2007generation,han2014quantum,szombati2016josephson}. However, the shortcoming with the devices, based on QDs, is the short range exchange interaction due to spatial separation in between them \cite{carvalho2003laser,szumniak2015long}.
Here, we introduce $\pi$-shaped two-qubit entangler device solely in solid state environment which connects two single qubit operations and generates Bell states. One dimensional nanowires instead of QDs have been proposed as an interconnect for entangling SQu because of long spin coherence length ($l_s$) \cite{frolov2013quantum,bercioux2015quantum,bandyopadhyay2002rashba}.

In this manuscript, we have described the quantum mechanical model for $\pi$-shaped quantum device. The $\pi$-shaped device consists of source, channel and output leads as shown in Fig.\ref{fig:device}.
Single SQu ($\left|\uparrow_s\right>$ or $\left|\downarrow_s\right>$) will be generated upon the application of electric field on sources S1 and S2  \cite{bringer2011spin}.
Then, these SQu are tunneled to channel by applying electric potential where they interact with SQu available within the channel ($\left|\uparrow_c\right>$ or $\left|\downarrow_c\right>$). In the channel, exchange interaction is responsible for two-qubit entanglement ($\left|\uparrow_s\right>$ or $\left|\downarrow_s\right>$)$\bigotimes$($\left|\uparrow_c\right>$ or $\left|\downarrow_c\right>$) (detailed description of device is provided in section II). For two-qubit entanglement, we have used analytical approach towards calculating the energy spectrum for SQu in source and channel. This energy spectrum provides available energy states for qubit to interact.
In section III, we have provided the generic equations for energies of single and two-qubit states to check further whether these equations have electrical tunability. We also calculated the energy spectrum in channel around Fermi level using Wentzel–Kramers–Brillouin (WKB) approximation along with iterative quasi linearization method (QLM) \cite{krivec2004quasilinear,mandelzweig1999quasilinearization}. This energy spectrum helps in determining the available energy levels for source SQu to interact with channel SQu. Exchange interaction between these SQu can be turned on/off by applying magnetic field along with electrical control over these energy states helping in creating Bell states.
These interaction pairs are then extracted to different parts of chip for QIP; due to entanglement between these pairs, measurement on one pair effects the other pair.
In section IV, we described how to implement the $U_{\sqrt{SWAP}}$ and CNOT from these electrically controlled Bell states. In section V, we proposed the scheme for physical implementation of $\pi$-shaped quantum device.

\section{$\pi$-shaped Quantum Device}
\begin{figure}[h!]\centering
    \includegraphics[scale=.8]{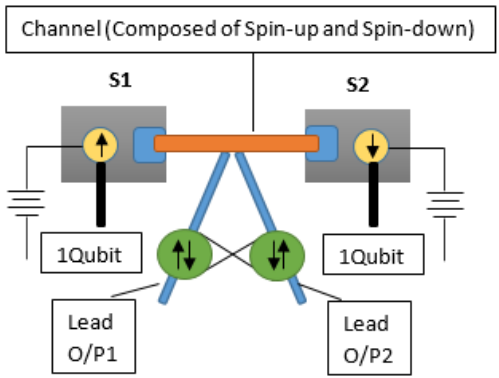}
    \caption{Schematic diagram for a 2-qubit device.}\label{fig:device}
\end{figure}

In Fig. \ref{fig:device}, S1 and S2 need to be highly spin polarized because upon application of electrical field, there is momentum dependent spin-splitting. For this purpose, the best suitable candidates are ceramic half-metals such as $La_{0.7}Sr_{0.3}MnO_3$ or $La_{0.7}Ca_{0.3}MnO_3$ that have spin-polarization of $\approx 100\%$ at $4.2$ K  \cite{chen2001spin, nadgorny2001origin}. Also, the dilute-magnetic-semiconductors (DMS) such as $GaMnAs$ are also the promising candidates as they show significant spin polarization even at $\approx$ $125K$ \cite{piano2011spin}. $CoFe_{2}O_4$ after carbon doping, showed $60\%$ spin polarization near room-temperature \cite{ramos2008influence, jeon2012thermal} and can also be used. These polarized spins are injected into the channel with electrical gating where channel has long `$l_s$' to avoid dephasing during interaction with source spins. For this purpose, $InAs$ can be used which shows `$l_s$' of 4.5 to 81 nm with nanowire diameter range of about 10 nm \cite{pan2014controlled}.

There are different interaction possibilities between source SQu and channel SQu  $\left(\left|\uparrow_s\right>\bigotimes\left|\uparrow_c\right>, \left|\uparrow_s\right>\bigotimes\left|\downarrow_c\right>, \left|\downarrow_s\right>\bigotimes\left|\uparrow_c\right>, \left|\downarrow_s\right>\bigotimes\left|\downarrow_c\right> \right)$; all these possibilities lie in superposition of spins within the channel due to exchange interaction. An ac magnetic field ($B_0(t)$) is applied across the channel to control spin orientation of electrons. The length of channel is less than `$l_s$' for the exchange interaction to be effective. A pair of these possibilities can be  extracted to output 1 (O/P1) and output 2 (O/P2) ($\pi$-shaped magnetic leads) by applying an appropriate electric potential. To extract these exchange pairs, magnetic or semiconductor (SC) o/p leads can be used. The use of SC contacts as o/p leads will serve the purpose of avoiding possibility of spin de-phasing \cite{hevroni2016suppression, higginbotham2015parity,nenashev2015quantum,brunner2011two}. Fabrication of thin channel can lead to interesting effects such as the Coulomb Blockade effect which may allow few electrons to enter into the channel thus, allowing efficient spin injection.

For the implementation and manipulation of 2-qubit entanglement among different interaction possibilities, the exchange coupling is turned on by taking electronic spins out of resonance under applied magnetic field $B_0(t)$. This can be done by applying additional magnetic field $B_0(t)+\partial B(t)$ across the channel. This exchange interaction between source and channel spins generates Bell states:
\begin{gather}
\left|\phi^\pm\right>=\frac{1}{\sqrt{2}}\bigl(\left|\uparrow_s\uparrow_c\right>\pm\left|\downarrow_s\downarrow_c\right>\bigr)\\
\left|\psi^\pm\right>=\frac{1}{\sqrt{2}}\bigl(\left|\uparrow_s\downarrow_c\right>\pm\left|\downarrow_s\uparrow_c\right>\bigr)
\end{gather}
We have formulated the energy spectrum for SQu in source and channel by constructing an electrically driven Hamiltonian in eqs. \ref{eq:1.9} and \ref{eq:2.20}. The corresponding energy states help in implementation of these Bell states. Divincenzo criteria requires that the quantum device should be able to implement 2 qubit (Bell state in our case) alongwith the 1 qubit (rotation operation) operations for universal quantum computation.
For the implementation of single qubit, $B_0(t)$ is brought into resonance with energy difference between source and channel spins. This resonance eliminates the exchange coupling between source spins and channel spins and thus, we have the single qubit operation:

\begin{gather}
\left|\psi\right>=\alpha(\left|\uparrow_s\uparrow_c\right>+\left|\downarrow_s\uparrow_c\right>)=\alpha(\left|\uparrow_s\right>+\left|\downarrow_s\right>)\otimes\left|\uparrow_c\right>\\
\left|\phi\right>=\alpha(\left|\uparrow_c\uparrow_s\right>+\left|\downarrow_c\uparrow_s\right>)=\alpha(\left|\uparrow_c\right>+\left|\downarrow_c\right>)\otimes\left|\uparrow_s\right>
\end{gather}
In subsection \ref{section:A}, we have formulated electrically driven energy states for single qubit. By adjusting the electrical control, single qubit can be implemented. Universal quantum computation can be performed by combined the $U_{\sqrt{SWAP}}$ gate and one qubit operation. In our case, the $U_{\sqrt{SWAP}}$ is implemented through adjusting the voltage or exchange  interaction between SQu in the channel. Single qubit is implemented through S1 and S2 by controlling different spin orientation of electrons as shown in Fig. \ref{fig:device}. We have also provided two other possible structures for $\pi$-shaped quantum device in supplementary file (S1) which is capable of implementing Bell states.

\section{Energies Spectrum for SQu in Source and Channel}
Our device is a combination of thin films (sources, S1 and S2) and nanowires (channel and o/p). In thin-film, electrons behave as 2-dimensional electron gas (2DEG) and in nanowires, electrons experience a quartic potential due to 2-dimensional confinement. And in both, electrons experience different columbic interaction. In following subsections (\ref{section:A} and \ref{section:B}), we have formulated the energy eigenstates/eigenvale for `$\uparrow$' and `$\downarrow$' electrons by setting electrical influenced Hamiltonian for 2-d and 1-d system. These electrical driven eigenstates/eigenvalue help in achieving Bell state and single qubit operation.

Energies related to SQu in source and channel are to be calculated using matrix mechanics approach by considering the Columbic and Rashba interactions between SQu. Energy spectrum in a channel around the Fermi-level is calculated using WKB approximation along with quasi linearization method (QLM) \cite{krivec2004quasilinear,mandelzweig1999quasilinearization}. Coupling of these SQu is explained by exchange Hamiltonian and the strength of the exchange constant $J_{ex}$ within the channel length is to be determined by the difference between lowest triplet energy and highest singlet energy in the channel \cite{hu2000hilbert}.
\subsection{Source:}\label{section:A}
In source, electrons behave as the 2-dimensional electron gas.  We assume a plane wave form ($\psi_k(y)$) of electrons behavior due to the translational invariance of the Hamiltonian along this axis:
\begin{equation}
\Psi_{n,k,\chi}(x,y) =\phi_n(x)\psi_k(y)
\left[{\begin{array}{c}
        \left|\uparrow\right> \\
        \left|\downarrow\right>
\end{array}}\right]
\end{equation}
We choose y-axis to be the propagation direction, the functions $\phi_n(x)\left|\uparrow\right>$ and $\phi_n(x)\left|\downarrow\right>$ represent spin-up and spin-down wave functions along the perpendicular axis to the propagation direction; spin-up is oriented along +z and spin-down along -z direction.
\begin{equation}\label{eq:1.8}
\Psi_{n,k,\chi}(x,y) =\sqrt{\frac{2}{L_x}}\sin\left({\frac{\pi x}{L_x}}\right)\exp[{\dot{\iota}k.y}]
\left[{\begin{array}{c}
        \left|\uparrow\right> \\
        \left|\downarrow\right>
\end{array}}\right]
\end{equation}
The total Hamiltonian consists of intrinsic Hamiltonian and the Rashba spin-orbit interaction (RSOI) potential. The intrinsic Hamiltonian includes the confinement potential due to harmonic behavior of electrons and interaction potential due to Coulombic interaction between electrons. The Coulombic interaction has taken a logarithmic form for 2DEG \cite{osada2013interacting,partoens2004structure}.
\begin{equation}\label{eq:1.9}
\begin{split}
\hat{H}=&\hat{H}_0+\hat{H}_{so}=\frac{\hat{P}_x^2}{2m^*}+\frac{\hat{P}_y^2}{2m^*}+\frac{1}{2}m^*\omega^2(\hat{x}.\hat{x}+\hat{y}.\hat{y})\\
&-\beta\ln\left[{\frac{x-y}{R}}\right]+\frac{\alpha}{\hbar}(\sigma_xP_y-\sigma_yP_x)
\end{split}
\end{equation}
where;
$\beta=\frac{q^2K}{R}$, $\alpha=\frac{e\hbar E}{2(m^*c)^2}$ and $m^*$ is the effective mass.
Energy eigenvalue and energy eigen vector is to be formulated by matrix mechanics as shown below:
\begin{equation}\label{eq:1.10}
\left<H\right> =
\left[{\begin{array}{cc}
     \left<H_{11}\right>&\left<H_{12}\right>\\
     \left<H_{21}\right>&\left<H_{22}\right>
\end{array}}\right]
\end{equation}
with help of Eqs. \ref{eq:1.8} and \ref{eq:1.9}, we can write:
\begin{equation}
\left({\begin{array}{cc}
     \left<H_{11}\right>\to & \left<\Psi_{n,k,\uparrow}(x,y)\right|\hat{H}\left|\Psi_{n,k,\uparrow}(x,y)\right>\\
     \left<H_{12}\right>\to & \left<\Psi_{n,k,\uparrow}(x,y)\right|\hat{H}\left|\Psi_{n,k,\downarrow}(x,y)\right>\\
     \left<H_{21}\right>\to &  \left<\Psi_{n,k,\downarrow}(x,y)\right|\hat{H}\left|\Psi_{n,k,\uparrow}(x,y)\right>\\
     \left<H_{22}\right>\to & \left<\Psi_{n,k,\downarrow}(x,y)\right|\hat{H}\left|\Psi_{n,k,\downarrow}(x,y)\right>
\end{array}}\right)
\end{equation}
We have provided the exact vales for `$H_{11}, H_{12}, H_{21}, H_{22}$' in supplementary file S1.
Energy difference between `$\uparrow$' and `$\downarrow$' is `$\Delta E$', is very useful for drifting electrons from source to channel.
\begin{equation}
\Delta E=E_{\downarrow}-E_{\uparrow}=\sqrt{(\left<H_{11}\right>-\left<H_{22}\right>)^2+4\left<H_{12}\right>\left<H_{21}\right>}
\end{equation}
\begin{figure}[h!]\centering
    \includegraphics[scale=.55]{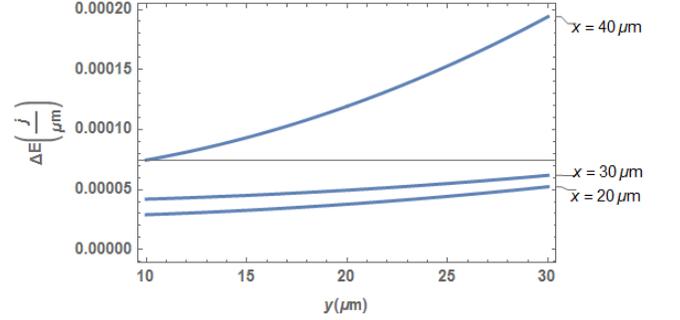}
    \caption{Change in `$\Delta E$' with respect to propagation direction of spin electrons at different value of width of source `X' (direction perpendicular to the propagation direction.)}\label{fig:deltaE}
\end{figure}
Figure \ref{fig:deltaE} shows the change in `$\Delta E$' with respect to propagation direction (-y) of spin electrons at different values of source width `x' (direction perpendicular to the propagation direction). Dependence of `$\Delta E$' on `x' is due to chosen x-dependent wave function. For practical purposes, it is very useful for the tunneling of spin-polarized electrons (either up or down). This tunneling can be controlled by applied electric-field which appears due to the Rashba term in Hamiltonian.
\subsection{Channel:}\label{section:B}
\subsubsection{Energy Spectrum in channel around fermi-level:}
SQu, drifted from source will interact with SQu available in different energy state in channel around fermi-level. This energy spectrum tells us about which energy states are available for interaction and how much energy of source SQu is needed for exchange interaction.
A complete wave function representing a particle in 1-dimension (let's say y-direction) under the influence of Rashba effect is represented below:
\begin{equation}
\Psi_{n,k_y,\sigma_y}(y)=\phi_n(y)\psi_{k_y}(y)\left|\sigma_y\right>
\end{equation}
where; $\phi_n(y)$ is the n-th eigenfunction of particle under the potential $V_y$(quartic potential in this case), and $\psi_{k_y}(y)$ is the shifted plane wave-function under the influence of Rashba interaction. The shift, determined by Pauli spin matrix ($\sigma_y$). $\phi_n(y)$ is to be determined by using WKB approximation along with quasi linearization method. The operation of Hamiltonian on a state ``$\phi_n(y)$'', is described as:
\begin{equation}\label{eq:1.12}
(\frac{-\hbar^2}{2m^*}\frac{\partial^2}{\partial y^2}+\frac{m^*\omega^2}{8a^2}(y^2-a^2)^2+V_{c}(y))\phi_n(y)=E_n\phi(y)
\end{equation}
For simplicity, take $\hbar=1$, $m=1$ and $m^*\neq 1$ (it depends on the periodic potential exhibited by lattice structure of materials), `$a=\sqrt{\hbar/(m\omega)}$' is harmonic length and by rearranging eq. \ref{eq:1.12}, we get:
\begin{equation}
\frac{\partial^2}{\partial y^2}\phi_n(y)+k_n^2(y)\phi_n(y)=0
\end{equation}
where; $k_n^2(y)=2m^*\left[E_n-\frac{m^*\omega^2}{8a^2}(y^2-a^2)^2-V_c(y)\right]$:
writing in terms of Riccati equation by substituting $\frac{\acute{\phi_n(y)}}{\phi_n(y)}=l(y)$, we get;
\begin{equation}\label{eq:1.1}
\frac{\partial l(y)}{\partial y}+k^2(y)+l^2(y)=0
\end{equation}
Applying QL to eq. \ref{eq:1.1} we get:
\begin{equation}
\frac{\partial l_n(y)}{\partial y}+2(l_{n-1}(y))l_n(y)=l^2_{n-1}(y)-k_n^2(y)\label{eq:1.2}
\end{equation}
For `n=1', eq. \ref{eq:1.2} becomes:
\begin{equation}
\frac{\partial l_1(y)}{\partial y}+2(l_{0}(y))l_1(y)=l^2_{0}(y)-k_n^2(y)\label{eq:1.3}
\end{equation}
The zero iterate should be based on physical considerations. Let us consider first an initial guess $l_0(y)=-g y$, where `g' is a nonzero constant which defines a harmonicity of the system. By comparing eq. \ref{eq:1.3} with linear differential equation:
\begin{equation}
\acute{l_1(y)}+P(y)l_1(y)=Q(y)\label{eq:1.4}
\end{equation}
Solution to eq. \ref{eq:1.4} is:
\begin{equation}
l_1(y)=\frac{1}{u(y)}\int_0^y u(s)\times Q(s)ds\label{eq:1.5}
\end{equation}
where: $u(y)=\exp(\int P(y)dy)$, $P(y)=2(-g y)$ and $Q(y)=(-g y)^2-2m^*\left[E_1-\frac{m^*\omega^2}{8a^2}(y^2-a^2)^2-V_c(y)\right]$. For determination of $E_1$, we make the following approximation:
`$l_1(y)\exp[\int(-2gy)dy] \to 0$' as $y\rightarrow\infty$ and upon simplifying, `$E_1$' becomes;
\begin{equation}\label{eq:1.19}
\begin{split}
&E_1=\frac{1}{2m^*\mathlarger{\int_0^\infty}\exp[\int(-2gs)ds] ds}\times\\
&\mathlarger{\int_0^\infty} \exp\left[\int(-2gs)ds\right]\times\left((-g s)^2+\frac{m^*\omega^2}{4a^2}(s^2-a^2)^2+2V_c(s)\right) ds
\end{split}
\end{equation}
Similarly, $E_2, E_3, ...$ can be determined by inserting `n=2,3,...' in eq.\ref{eq:1.2} and following the iterations. The $E_1, E_2, E_3, ...$ are available energy states for ($\left|\uparrow_s\right>$ or $\left|\downarrow_s\right>$) to interact with ($\left|\uparrow_c\right>$ or $\left|\downarrow_c\right>$).
\subsubsection{Dynamics for 2 qubit and exchange interaction}
In the channel, intrinsic potential consists of harmonic and Coulombic potential. Since, the channel is 1-dimensional (spin electrons are propagating along y-direction), there is a strong parabolic confinement along `x-' and `y-' directions. So, quartic harmonic potential is chosen due to merging of two harmonic potentials \cite{burkard1999coupled,ceausu2014phase}. And the effective 1-dimensional Coulombic interaction between electrons is an appropriate choice for the channel \cite{bednarek2003effective}. Extrinsic potential consists of Rashba term having momentum only along y-direction. Hence, the total Hamiltonian becomes:
\begin{equation}\label{eq:2.20}
\hat{H}=\frac{\hat{P}_y^2}{2m^*}+\frac{m^*\omega^2}{8a_B^2}(x^2-a^2)^2+V_c(y)+\frac{\alpha}{\hbar}(\sigma_xP_y)
\end{equation}
where, `$a_B=\sqrt{\frac{\hbar}{m\omega}}$' is harmonic length and $V_c(y)$ is;

\begin{equation}
\begin{split}
V_c(y)&=\sqrt{\frac{\pi}{2}}\frac{k}{l}Erfcy\left(\frac{y}{\sqrt{2}}\right)\\
&=\sqrt{\frac{\pi}{2}}\frac{k}{l}\exp{\left[\frac{y^2}{2l^2}\right]}Erfc\left(\frac{y}{\sqrt{2}}\right)
\end{split}
\end{equation}
where: $k=\frac{q^2}{4\pi\varepsilon_0\varepsilon}$ and `$l$' is fermi-length for electrons.\\
A complete spin dependent wave function `$\Psi_{n,k,\chi}(x,y)$' is a product of particle state function, plane wave function (due to splitting) and spin selection.
\begin{equation}
\begin{split}
&\Psi_{n,k,\chi}(x,y)=\\
&\left(\frac{1}{a_B^2\lambda^2\pi^2}\right)^{\frac{1}{4}}\exp\left[{-\frac{y^2}{2\lambda^2}-\frac{x^2}{2a_B^2}}\right]\exp[{\dot{\iota}{k}.{x}}]
\left[{\begin{array}{c}
        {\uparrow\uparrow} \\
        {\uparrow\downarrow} \\
        {\downarrow\uparrow} \\
        {\downarrow\downarrow}
\end{array}}\right]
\end{split}
\end{equation}
$4\times4$ matrix can be constructed by taking summation on $\left<\Psi_{n^*,k^*,\chi_{ji}}(x,y)\right|\hat{H}\left|\Psi_{n,k,\chi_{ij}}(x,y)\right>$.
\begin{equation}\label{eq:1.15}
\left<\hat{H}\right> =
\left[{\begin{array}{cccc}
      \left<H_0\right>  &  0                &  \left<H_R\right>         & 0               \\
      \left<H_R\right>  &  0                &   \left<H_0\right>        & 0                \\
        0               & \left<H_0\right>  &  0                        & \left<H_R\right>       \\
      0                 &\left<H_R\right>   &  0                        & \left<H_0\right>
\end{array}}\right]
\end{equation}
As a consequence, the eigen system for above matrix \ref{eq:1.15} becomes:

\begin{equation}
\begin{rcases}
  E_{\uparrow\uparrow}&=\left<H_0\right>-\left<H_R\right>\\
  \phi_{\uparrow\uparrow}&=\{1,-1,-1,1\}
\end{rcases}
\text{ $\left|\uparrow\uparrow\right>$}
\end{equation}
\begin{equation}
\begin{rcases}
  E_{\uparrow\downarrow}&=\left<H_0\right>+\left<H_R\right>\\
  \phi_{\uparrow\downarrow}&=\{1,1,1,1\}
\end{rcases}
\text{ $\left|\uparrow\downarrow\right>$}
\end{equation}
\begin{equation}
\begin{rcases}
  E_{\downarrow\uparrow}&=-\sqrt{\left<H_0\right>^2-\left<H_R\right>^2}\\
  \phi_{\downarrow\uparrow}(x,y)&=\{-1,-\frac{\left<H_0\right>+\sqrt{\left<H_0\right>^2-\left<H_R\right>^2}}{\left<H_R\right>},\\ &-\frac{-\left<H_0\right>^2+\left<H_R\right>^2-\left<H_0\right>\sqrt{\left<H_0\right>^2-\left<H_R\right>^2}}{\left<H_R\right>\sqrt{\left<H_0\right>^2-\left<H_R\right>^2}},1\}
\end{rcases}
\text{ $\left|\downarrow\uparrow\right>$}
\end{equation}

\begin{equation}
\begin{rcases}
  E_{\downarrow\downarrow}&=\sqrt{\left<H_0\right>^2-\left<H_R\right>^2}\\
  \phi_{\downarrow\downarrow}(x,y)&=\{-1,-\frac{\left<H_0\right>-\sqrt{\left<H_0\right>^2-\left<H_R\right>^2}}{\left<H_R\right>},\\ &-\frac{\left<H_0\right>^2-\left<H_R\right>^2-\left<H_0\right>\sqrt{\left<H_0\right>^2-\left<H_R\right>^2}}{\left<H_R\right>\sqrt{\left<H_0\right>^2-\left<H_R\right>^2}},1\}
\end{rcases}
\text{ $\left|\downarrow\downarrow\right>$}
\end{equation}

 where: $\left<H_0\right>$ and $\left<H_R\right>$ looks like:
\begin{equation}
\begin{split}
\left<H_0\right>=&\left<\left(\frac{1}{a_B^2\lambda^2\pi^2}\right)^{\frac{1}{4}}\exp\left[{-\frac{y^2}{2\lambda^2}-\frac{x^2}{2a_B^2}}\right]\exp[{-\dot{\iota}{k}.{x}}]\right|\\
&\hat{H_0}\left|\left(\frac{1}{a_B^2\lambda^2\pi^2}\right)^{\frac{1}{4}}\exp\left[{-\frac{y^2}{2\lambda^2}-\frac{x^2}{2a_B^2}}\right]\exp[{\dot{\iota}{k}.{x}}]\right>
\end{split}
\end{equation}

\begin{equation}
\begin{split}
&\left<H_R\right>=\\
&-\alpha\dot{\iota}\hbar\left<\left(\frac{1}{a_B^2\lambda^2\pi^2}\right)^{\frac{1}{4}}\exp\left[{-\frac{y^2}{2\lambda^2}-\frac{x^2}{2a_B^2}}\right]\exp[{-\dot{\iota}{k}.{x}}]\right|\\
&\hat{\frac{\partial}{\partial y}}\left|\left(\frac{1}{a_B^2\lambda^2\pi^2}\right)^{\frac{1}{4}}\exp\left[{-\frac{y^2}{2\lambda^2}-\frac{x^2}{2a_B^2}}\right]\exp[{\dot{\iota}{k}.{x}}]\right>
\end{split}
\end{equation}
These energy eigenstates ($E_{\uparrow\uparrow},E_{\uparrow\downarrow},E_{\downarrow\uparrow},E_{\downarrow\downarrow}$) lie in superposition due to exchange interaction between source and channel spin-electrons:
\begin{equation}\label{eq:2.32}
{H_{ex}}= J(t){\bf S_s.S_c}
\end{equation}
where, $J(t)$ is the time-dependent exchange constant. This exchange coupling can be turned on by applying magnetic field which takes electrons out of resonance and all these energy states will be entangled. It is also termed as the background current and can be calculated by calculating difference between lowest triplet energy and highest singlet energy states \cite{hu2000hilbert}.
Electrical control over all these energy states due to Rashba term ($\left<H_R\right>$) will give rise to desired outputs; $\left|\phi^\pm\right>$ and $\left|\psi^\pm\right>$ (Bell entangled states).

\section{Bell states and SWAP gate}
$U_{SWAP}$ operation can be implemented through Bell states, i.e.
\begin{equation}
\begin{split}
&(U_{SWAP})^\alpha=\\
&\left|\phi^+\right>\left<\phi^+\right|+\left|\phi^-\right>\left<\phi^-\right|+\left|\psi^+\right>\left<\psi^+\right|+e^{\dot{\iota}\alpha}\left|\psi^-\right>\left<\psi^-\right|
\end{split}
\end{equation}
In matrix form,
\begin{equation}\label{eq:2.34}
(U_{SWAP})^\alpha =
\left({\begin{array}{cccc}
      1  &  0                                     &  0                                           & 0               \\
      0  &  \frac{1+e^{\dot{\iota}\alpha}}{2}  &  \frac{1-e^{\dot{\iota}\alpha}}{2}        & 0                \\
      0  &  \frac{1-e^{\dot{\iota}\alpha}}{2}  &  \frac{1+e^{\dot{\iota}\alpha}}{2}        & 0                 \\
      0  &  0                                     &  0                                           & 1
\end{array}}\right)
\end{equation}
where; `$\alpha$' defines the nature of $U_{SWAP}$ operation and is determined by the time-evolution of exchange interaction of eq. \ref{eq:2.32}, that is:
\begin{equation}\label{eq:2.36}
{U_{s,c}}=\exp{\left[\frac{-\dot{\iota}}{\hbar}{\bf S_s.S_c}\int J(t)dt\right]}=\exp{\left[\frac{-\dot{\iota}}{\hbar}{\bf S_s.S_c}\alpha\right]}
\end{equation}
where: ${\bf S_s.S_c}=(\sigma_x\otimes\sigma_x+\sigma_y\otimes\sigma_y+\sigma_z\otimes\sigma_z)$, writing in terms of $U_{SWAP}=diag(1,\sigma_x,1)$:
\begin{equation}\label{eq:2.37}
{\bf S_s.S_c}=\frac{1}{4}\left(2U_{SWAP}-I\right)
\end{equation}
Inserting eq. \ref{eq:2.37} in eq. \ref{eq:2.36} and after simplifying we get:
\begin{equation}\label{eq:3.37}
\begin{split}
{U_{(s,c)}}=&\exp{\left[\frac{-\dot{\iota}}{4}\left(2U_{SWAP}-I\right)\alpha\right]}\\
&\exp{\left[\frac{\dot{\iota}\alpha}{4}\right]}\exp{\left[\frac{-\dot{\iota}\alpha U_{SWAP}}{2}\right]}
\end{split}
\end{equation}
From eq. \ref{eq:3.37}, the nature of the $U_{SWAP}$ gate can be adjusted by controlling the time evolution of exchange interaction.  By setting $\alpha=\frac{1}{\hbar}\int_{t1}^{t2} J(t)dt=\pi$ in eq. \ref{eq:2.34}, we get the $U_{SWAP}$ gate:
\begin{equation}
U_{SWAP} =
\left({\begin{array}{cccc}
      1  &  0  &  0        & 0               \\
      0  &  0  &  1        & 0                \\
      0  &  1  &  0        & 0                 \\
      0  &  0  &  0        & 1
\end{array}}\right)
\end{equation}
where, $t_1$ and $t_2$ are the gate operational time, during this time interval electrons should remain in gate region. By setting $\alpha=\frac{1}{\hbar}\int_{t1}^{t2} J(t)dt=\frac{\pi}{2}$, we get the $U_{\sqrt{SWAP}}$ gate;
\begin{equation}
U_{\sqrt{SWAP}} =
\left({\begin{array}{cccc}
      1  &  0                                     &  0                                           & 0               \\
      0  &  \frac{1+\dot{\iota}}{2}  &  \frac{1-\dot{\iota}}{2}        & 0                \\
      0  &  \frac{1-\dot{\iota}}{2}  &  \frac{1+\dot{\iota}}{2}        & 0                 \\
      0  &  0                                     &  0                                           & 1
\end{array}}\right)
\end{equation}
Now, it is known that one CNOT gate can be realized by two $U_{(SWAP)^{1/2}}$ gates and single-qubit unitary gate hence, six $U_{(SWAP)^{1/2}}$ gates are sufficient to implement any two-qubit operation \cite{fan2005optimal}.

\section{Fabrication Scheme for $\pi$-shaped Quantum Device}
For the fabrication of $\pi$-shaped quantum device, firstly nanowires are deposited in a random orientation after which they are brought together to form o/p leads by using a  scanning electron microscopy system equipped with nanomanipulator, e.g. the Raith eLINE Plus system. Incidently, the nanomanipulator may also be used for in-situ probing and electrical measurement at later stages. Subsequently, spray coating of photoresist may be performed to cover the nanowires without perturbing their location. Electron beam lithography (EBL) can be used to define the channel. Physical vapor deposition may be used to deposit the channel material for achieving conformal coverage. After liftoff, the same EBL process followed by metallization strategy defined previously, can be used to pattern the sources (S1 and S2) with dimension of approximately $\approx40\mu m\times40\mu m$. However the metallization, this time, will be of a magnetic metal. As per design, the leads are to be around 10-20 nm in diameter with channel length of 20 nm or less. These dimensions can easily be accessed by EBL. Proposed design in supplementary file  can be made using the same strategy, however, QDs of $\approx6$ nm may either be patterned by EBL or placed by nanomanipulator. The magnetic contacts can be made using magnetron sputtering and the fabrication of the top gate can be done using atomic layer deposition system. In order to protect the device, chemical etching is suggested by using suitable etchants such as HCl, HF, etc. Lift-off can be done using Acetone followed by washing in de-ionised water.

\section{Conclusion}
For multi-qubits quantum computing, entanglement needed to be preserved for QIP in intra- and inter-chips system. To acheive this landmark, we need a reliable solid state interconnect. In our manuscript, we proposed a practically viable $\pi$-shaped quantum device which is capable of entangling 2-qubits along with single qubit operations in solid state environment. We also presented the theoretical scheme to describe the dynamics of the  $\pi$-shaped quantum device. The novelty of this quantum device is that, it generates entanglement between 2 qubits in Bell basis which are maximally entangled states.

In describing the quantum dynamics of the device, we tunneled the source SQu to channel under an appropriate electric field where they experience exchange interaction $\left(\left|\uparrow_s\uparrow_c\right>, \left|\uparrow_s\downarrow_c\right>, \left|\downarrow_s\uparrow_c\right>, \left|\downarrow_s\downarrow_c\right> \right)$. As this exchange interaction is electrically controllable hence, gives rise to Bell states ($\left|\phi^\pm\right>$, $\left|\psi^\pm\right>$).
The $U_{\sqrt{SWAP}}$ and CNOT operations can be implemented using these Bell states for universal quantum computation.

\section{Acknowledgment}
The authors are thankful to the Higher Education Commission (HEC) of Pakistan for funding this research activity under Project No. 6040/Federal/NRPU/R$\&$D/HEC/2016.

\bibliographystyle{apsrev4-1}
\bibliography{jobname}

\end{document}


\title{Supplementary (S1) for the paper titled:\\A $\pi$-shaped Quantum Device for Implementation of Bell States in Solid State Environment}
\author{Aman Ullah}
 \affiliation{Department of Physics, School of Natural Sciences (SNS), National University of Sciences and Technology (NUST), Islamabad 44000, Pakistan}
 \author{Mohammad Ali Mohammad}
\affiliation{School of Chemical and Materials Engineering (SCME), National University of Sciences and Technology, Islamabad 44000, Pakistan}
\author{Mahmood Irtiza Hussain }
\affiliation{Institute for Quantum optics and Quantum Information
Technikerstr. 21a
6020 Innsbruck, Austria}
\author{Syed Rizwan}
\email{Corresponding author: Syed Rizwan\\ Email: syedrizwanh83@gmail.com}
\affiliation{Department of Physics, School of Natural Sciences, National University of Sciences and Technology (NUST), Islamabad 44000, Pakistan}

\date{\today}
\maketitle

\section{A $\pi$-shaped quantum structures for the implementation of Bell's state}
\begin{figure}[h!]\centering
    \includegraphics[scale=.3]{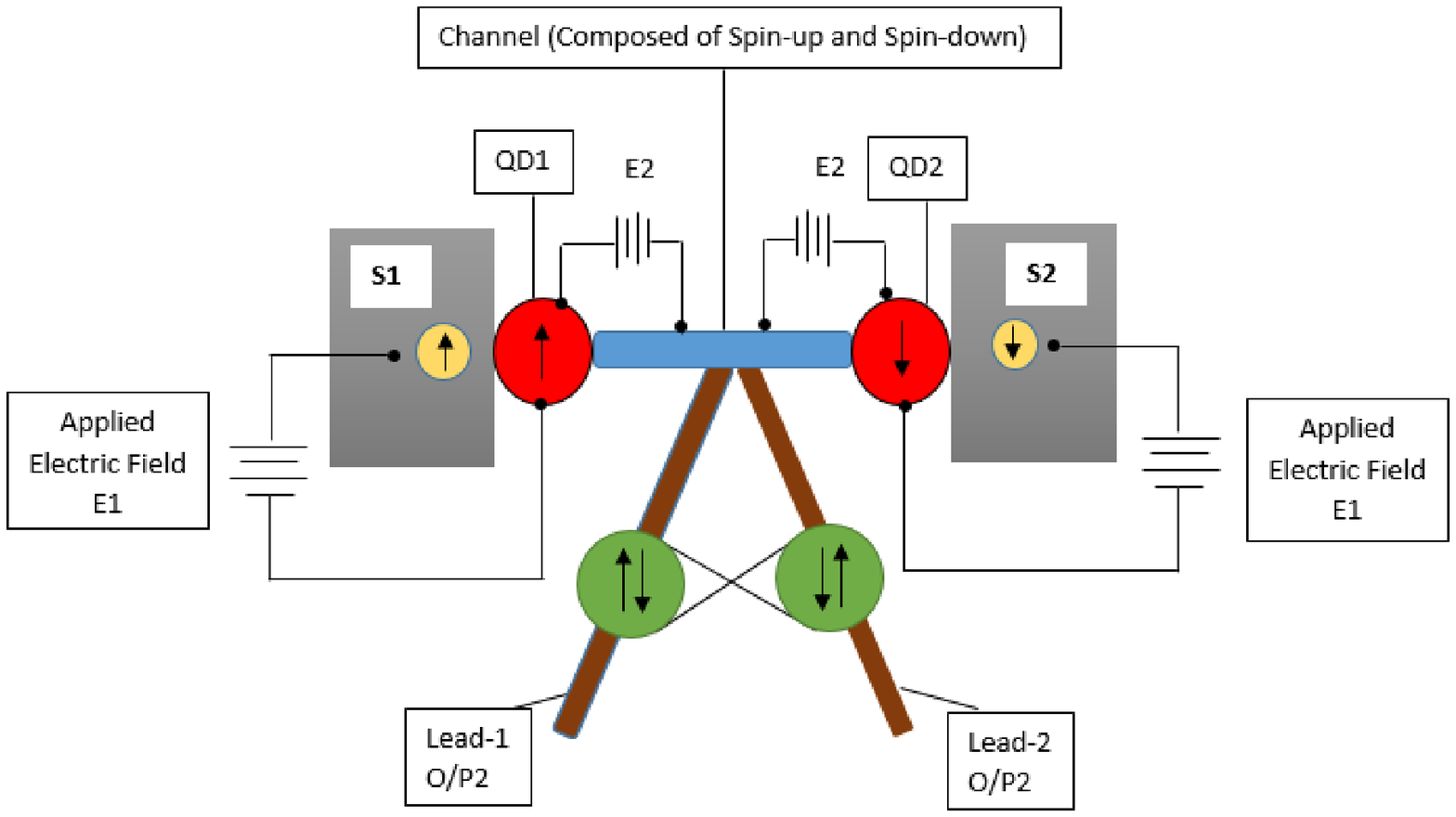}
    \caption{Circuit diagram for a 2-qubit device, second proposed implementation.}\label{fig:2qubit}
\end{figure}

Fig. \ref{fig:2qubit} shows the second proposed device where a Schottky barrier is created by introducing a quantum dot. In this device, electrons of particular spins tunnel to the channel and these dots are also responsible for 1 qubit rotation. Electrons are tunneled to the channel by biasing the dot.\\

\begin{figure}[h!]
    \includegraphics[scale=.3]{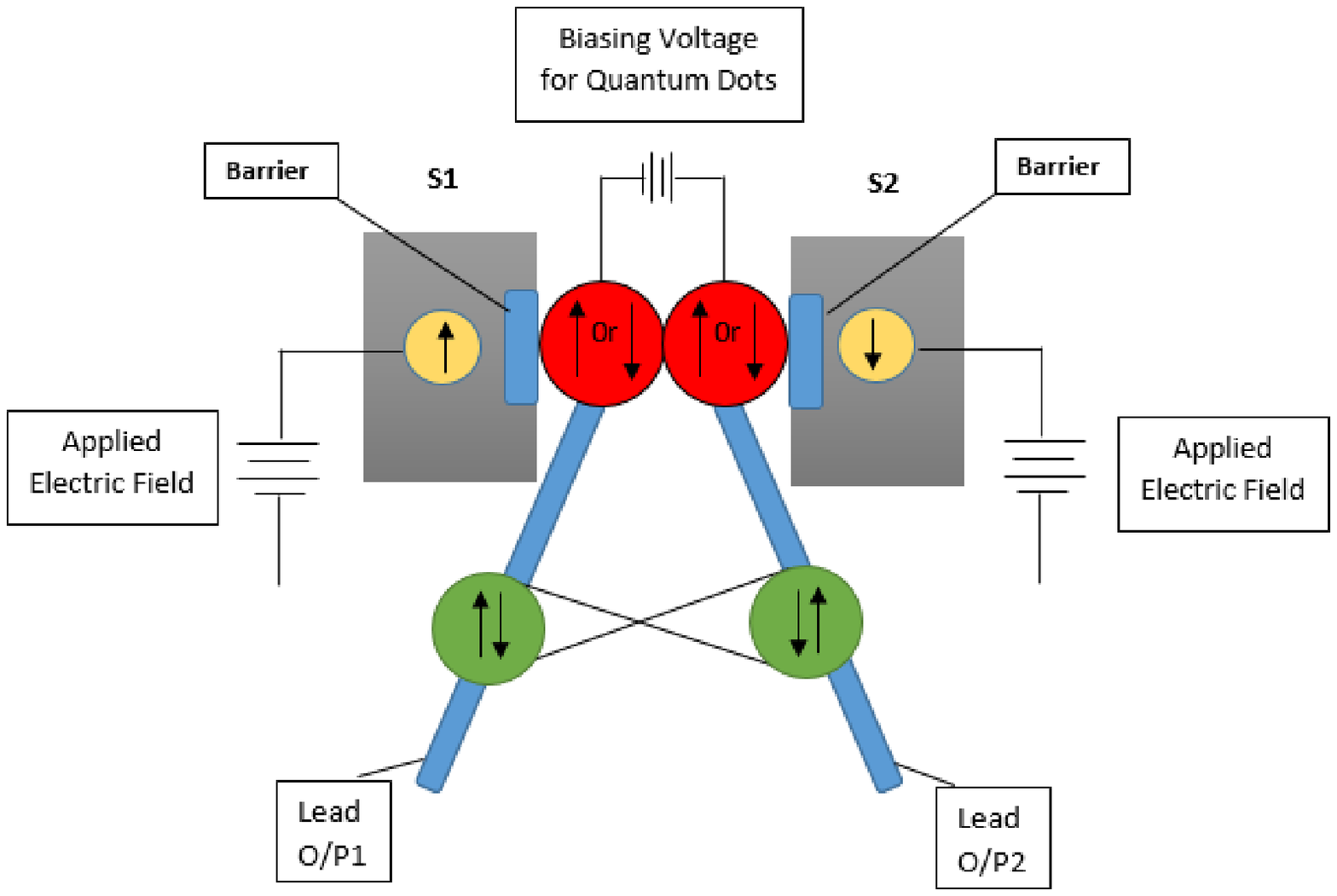}
    \caption{Circuit diagram for a 2-qubit device, third proposed implementation.}\label{fig:2qubit2}
\end{figure}

Fig. \ref{fig:2qubit2} shows the third proposed device where the channel is omitted and a pair of quantum dots operate as a channel. From these dots, electrons of different spins are extracted to the output leads. And information sharing process is the same as in the previous proposed device.\\

\section{Energy Eigenstates/Eigenvalues}
\subsection{For `$\uparrow$' and `$\downarrow$'}
\begin{equation}\label{eq:2.5}
\begin{rcases}
E_{\uparrow}=\frac{\left<H_{11}\right>+\left<H_{22}\right>-\sqrt{\left<H_{11}^2\right>-2\left<H_{11}\right>\left<H_{22}\right>+4\left<H_{12}\right>\left<H_{21}\right>}}{2}\\
\phi_\uparrow(x,y)=\{-\frac{-\left<H_{11}\right>+\left<H_{22}\right>+\sqrt{\left<H_{11}^2\right>-2\left<H_{11}\right>\left<H_{22}\right>+4\left<H_{12}\right>\left<H_{21}\right>}}{2\left<H_{21}\right>},1\}
\end{rcases}
\text{ $\left|\uparrow\right>$}
\end{equation}

\begin{equation}\label{eq:2.6}
\begin{rcases}
E_{\downarrow}=\frac{\left<H_{11}\right>+\left<H_{22}\right>+\sqrt{\left<H_{11}^2\right>-2\left<H_{11}\right>\left<H_{22}\right>+4\left<H_{12}\right>\left<H_{21}\right>}}{2}\\
\phi_\downarrow(x,y)=\{-\frac{-\left<H_{11}\right>+\left<H_{22}\right>-\sqrt{\left<H_{11}^2\right>-2\left<H_{11}\right>\left<H_{22}\right>+4\left<H_{12}\right>\left<H_{21}\right>}}{2\left<H_{21}\right>},1\}
\end{rcases}
\text{ $\left|\downarrow\right>$}
\end{equation}